\documentclass[fleqn,usenatbib]{mnras}

\usepackage{subfigure}
\usepackage{bm}
\newcommand{\feh} {\mbox{\rm [Fe/H]}}

\newcommand{\afe} {\mbox{\rm [$\alpha$/Fe]}}

\usepackage{newtxtext,newtxmath}

\usepackage[T1]{fontenc}

\DeclareRobustCommand{\VAN}[3]{#2}
\let\VANthebibliography\thebibliography
\def\thebibliography{\DeclareRobustCommand{\VAN}[3]{##3}\VANthebibliography}

\usepackage{graphicx}	
\usepackage{amsmath}	

\title[The role of radial migration on tracing lithium evolution in the Galactic disk]{The role of radial migration on tracing lithium evolution in the Galactic disk}

\author[H. P. Zhang et al.]{
Haopeng Zhang,$^{1}$
Yuqin Chen,$^{1,2}$\thanks{E-mail: cyq@nao.cas.cn}
Gang Zhao,$^{1}$
Shaolan Bi,$^{2}$
Xianfei Zhang$^{2}$
and Xiangxiang, Xue$^{1,2}$
\\
$^{1}$CAS Key Laboratory of Optical Astronomy, National Astronomical Observatories, Beijing 100101, China\\
$^{2}$Institute for Frontiers in Astronomy and Astrophysics, Beijing Normal University,  Beijing 102206, China\\
}

\date{Accepted XXX. Received YYY; in original form ZZZ}

\pubyear{2022}

\begin{document}
\label{firstpage}
\pagerange{\pageref{firstpage}--\pageref{lastpage}}
\maketitle

\begin{abstract}
With the calculated guiding center radius $R_{guiding}$ and birth radius $R_{birth}$, we investigate the role of radial migration on the description of lithium evolution in the Galactic disk based on the upper envelope of the A(Li) vs. $\feh$ diagram. 
Using migration distances, we find that stars in the solar neighborhood are born at different locations in the galactic disk, and cannot all be explained by models of chemical evolution in the solar neighborhood. 
It is found that the upper envelope of the A(Li) vs. $\feh$ diagram varies significantly with $R_{birth}$, which explains the decrease of Li for super-metal-rich (SMR) stars because they are non-young stars born in the inner disk. 
The upper envelope of Li-$R_{birth}$ plane fits very well with chemical evolution models by Grisoni et al. for $R_{birth} = 7 - 12$ kpc, outside which young stars generally lack sufficient time to migrate to the solar neighborhood.
For stars born in the solar neighborhood, the young open clusters and the upper envelope of field stars with age $<$ 3 Gyr fit well with theoretical prediction.
We find that calculations using stars with ages less than 3 Gyr are necessary to obtain an undepleted Li upper envelope, and that stars with solar age (around 4.5 Gyr) have depleted around 0.3 dex from the original value based on the chemical evolution model of Grisoni et al.
\end{abstract}

\begin{keywords}
stars: abundances -- Galaxy: abundances -- Galaxy: disc -- Galaxy: evolution
\end{keywords}



\section{Introduction}

The evolution of the lithium abundance (in this paper, all refer to the main isotope of lithium, $^7$Li) has still been a controversial issue of astrophysics so far.
An important issue is its primordial abundance: the initial lithium abundance predicted by the standard BB Nucleosynthesis (SBBN) model is A(Li) $\sim$ 2.7 dex \citep{pitrou2018precision}, while the most metal-poor halo dwarf stars in the Milky Way and debris from  the Gaia-Sausage-Enceladus galaxy give a so-called "Spite plateau" of A(Li) $\sim$ 2.2 dex \citep{spite1982abundance,bonifacio1997primordial,shi2007lithium,simpson2021galah,zhao2021low}.
For disk stars, lithium increases with metallicity as shown in \cite{chen2001lithium} based on high resolution spectra obtained at Xinglong station \citep{zhao2001coude}, and the upper envelope of Li-$\feh$ diagram is used to trace lithium evolution. 
However, with the continuous data release of large spectroscopic surveys, such as Gaia-ESO \citep{gilmore2012gaia,randich2013gaia}, GALAH \citep{de2015galah}, LAMOST \citep{cui2012large,deng2012lamost,zhao2012lamost}, etc., some works identified a decrease in lithium abundance at the super-solar metallicity  \citep[e.g.][]{mena2015li,guiglion2016ambre,bensby2018exploring,fu2018gaia,stonkute2020high}, which is an anomalous phenomenon.
In order to explain this phenomenon, \cite{prantzos2017ambre} reduced the lithium yields at the super-solar metallicity, while others suggested that it is due to the selection effect caused by old stars with lithium depletion\citep[e.g.][]{randich2020gaia,charbonnel2021behaviour}.
In addition, lithium in planet-host stars shows a lower value than non-planet-host stars \citep[e.g.][]{chen2006comparative}. 
Since many planet-host stars are super-metal-rich (SMR) stars, it is possible that the presence of planet also affect Li abundances.
\cite{guiglion2019explaining} suggested that this decrease is related to the radial migration: older stars born in the inner regions of the galactic disk migrated to the solar vicinity, which explains the presence of SMR stars in the solar neighborhood and makes the upper envelope of A(Li) vs. $\feh$ diagram invalid for tracing lithium evolution at the metal-rich end.
Recently, \cite{dantas2022gaia} analysed the SMR stars from the sixth internal data release (iDR6) of the Gaia-ESO Spectroscopic Survey (GES) and found that the older they are, the more Li depleted they are, and suggested that they migrated from the inner Galaxy based on their chemo-dynamic features.

The radial migration has been demonstrated to play an important role in the chemodynamical evolution of the galactic disk, both observationally \citep[e.g.][]{yu2012test,haywood2013age,bovy2016stellar} and theoretically \citep[e.g.][]{sellwood2002radial,minchev2011radial,minchev2013chemodynamical}.
There are two main mechanisms that can cause stars to stray away from their birth radius: blurring and radial migration (churning).
Blurring is the epicyclic motion of stars from their guiding centers, with the orbit becoming heated over time, but with the angular momentum conserved.
Radial migration (churning) is a change in the angular momentum of stars due to resonance interactions with non-axisymmetric structure, such as transient spiral arms \citep[e.g.][]{sellwood2002radial,Roskar2008riding} and the resonance overlap between the bar and spiral arms \citep[e.g.][]{minchev2010new,kubryk2013radial}, or non-resonance interactions, such as minor mergers \citep[e.g.][]{quillen2009radial,minchev2014new}.

Recovering the birth radius of stars is an important way to study the radial migration and the enrichment history of the Milky Way.
\cite{minchev2018estimating} presented a semi-empirical, largely model-independent approach for estimating the birth radius of stars using their ages and metallicities, based on an assumption for the Interstellar Medium (ISM) metallicity distribution in the disk.
Later, using simulations from the NIHAO-UHD project, \cite{lu2022reliability} found that the above method is reliable for inferring precise stellar birth radii from the start of stellar disk formation 10 Gyr ago to the present.

Although \cite{guiglion2019explaining} and \cite{dantas2022gaia} had attributed the decrease in lithium abundance at the super-solar metallicities to radial migration, 
analysis of the lithium abundance envelope by quantifying the migration distance is still lacking.
\cite{minchev2019yule} proposed that studying the relationship between lithium abundance and birth radius could provide stronger constraints on chemical evolution models.
In this paper, a first attempt is made to quantify the effect of radial migration on the evolution of lithium abundance in the Galactic disk by calculating the birth positions and migration distances of the stars.
In Sect.~2, we describe the data collection, processing, and calculation of the birth radius $R_{birth}$ and radial migration distance of stars.
Sect.~3 explains the effect of radial migration on the maximum lithium abundance envelope by comparison with the theoretical model gradient.
Finally, the results are summarized in Sect.~4.

\section{Data and Methods}

We used the lithium abundance catalog of \cite{romano2021gaia}, which was based on the GES iDR6.
This catalog provided a homogeneous sample of 26 open clusters with undepleted lithium abundances and 3210 field stars, covering large ranges of ages and galactocentric distances.
The lithium abundances and stellar parameters of open cluster members and field stars were homogeneously obtained from observations of the multi-subject optical fiber facility FLAMES \citep[Fibre Large Array Multi Element Spectrograph;][]{pasquini2002installation}.
The lithium abundances of the clusters represent the average maximum lithium abundance of the member stars.
This work did not carry out non-local thermal equilibrium (NLTE) corrections of the lithium abundance because it can be negligible for the majority of stars in the sample \citep[e.g.][]{zhao2016systematic}.
In addition, lithium can be depleted and occasionally enriched in giant stars \citep[e.g.][]{yan2018nature,yan2021most,kumar2020discovery}, and thus only dwarf stars are included in the sample.
We removed stars with undetermined lithium abundances and upper limits.
The remaining number of field stars is 2184 with the log $g$ range of $3.6-4.6$ dex, and a effective temperature ($T_{eff}$) range of $5300-7000$ K.
The choice of 5300 K aims to avoid the significant lithium destruction for the coolest dwarf stars.

Based on the kinematic parameters given by \cite{romano2021gaia}, we calculate the guiding center radius $R_{guiding}$ for open clusters and field stars using $galpy$ \citep{bovy2015galpy}.
Then, we calculate the stellar birth radius ($R_{birth}$) in the same approach as \cite{chen2020open,zhang2021radial}, i.e. based on the relation of $\feh_{star} - \feh_{ISM}(R_{\odot}, t) = gradient_{\feh} (R_{birth} - R_{\odot})$, where $\feh_{star}$ is $\feh$ of the star and t is the age of the star, and $R_{\odot}$ represents the solar radius of 8 kpc \citep{reid1993distance,bovy2012milky}. 
The evolution of ISM metallicity gradient ($gradient_{\feh}$) and ISM metallicity at the solar radius ($\feh_{ISM}(R_{\odot}, t)$) are taken from \cite{minchev2018estimating} and age information is from \cite{romano2021gaia}.
Because of the selection criteria of the GES, the sample does not include field stars younger than 1 Gyr.
To ensure that there is no systematic bias in the comparison with the theoretical model presented in Sect.~3, we shifted the metallicity of the ISM at the solar radius given by \cite{minchev2018estimating} up by 0.05 dex in our subsequent analysis.
As the results of \cite{minchev2018estimating} are based on observations, this is a good alternative in the lack of direct observations of ISM metallicity
\citep[see the discussion in][for details on the  reliability of ISM metallicity]{zhang2021radial}.
We regard $R_{guiding} - R_{birth}$ as the migration distance because radial migration (churning) can change the $R_{guiding}$, whereas blurring does not, and $R_{guiding}$ is a good proxy for the current radial position of the star \citep{chen2020open}.

\section{Results}

\subsection{The upper envelope in the A(Li) vs. $\feh$ diagram}
The upper panel of Fig.~1 shows the lithium abundances as a function of $\feh$ for the open clusters and field stars in the solar neighbourhood (7 kpc $< R_{gc} <$ 9 kpc) of the sample. 
To analyse the upper envelope, we used the same method as \cite{lambert2004lithium,guiglion2016ambre,guiglion2019explaining}: we binned the entire sample of field stars in $\feh$, and then calculated the average lithium abundance of the six stars with the highest lithium abundance within each bin. 
This line is also plotted on the upper panel of Fig.~1, with the error bars indicating the standard deviation of the lithium abundance of the six stars mentioned above in each $\feh$ bin.
The lithium envelope rises from A(Li) $=$ 2.0 dex at $\feh =$ -0.9 dex to A(Li) $=$ 3.3 dex at $\feh =$ 0.0 dex, after which it decreases to A(Li) $=$ 2.8 dex at $\feh =$ 0.3 dex.
Also, we identified high- and low-$\alpha$ stars in the same criteria as \cite{guiglion2016ambre,guiglion2019explaining,romano2021gaia} in order to make the results more comparable, but did not further group them according to the metallicity, with the aim of visualizing the variation in the lithium abundance of field stars throughout the range of metallicity after excluding the effects of $\alpha$ elemental abundances. 
A fraction of the stars lack $\afe$ measurements, eventually 337 and 574 high- and low-$\alpha$ stars are available respectively.
We obtained lithium envelopes for each subsample in the solar neighborhood using the same approach as above, and the results are shown in the lower panel of Fig.~1. 
The trend of the lithium envelope with metallicity in each subsample is similar to the upper panel, with a decreasing trend at the metal-rich side. 
The transition between the thick and the thin disk in the lower panel seems to explain the dip of the envelope at $\feh =$ -0.3 dex in the upper panel.
However, the upper envelope of A(Li) vs. $\feh$ diagram for the thin disk at $\feh =$ 0.0 dex is 3.0 dex, but one star has the highest A(Li) of 3.3 dex. 
Since the lower panel has a smaller sample than the upper one,  this discrepancy is due to the lack of stars with high Li abundance in the lower panel.
Therefore, the upper panel is more suitable to trace the evolution of lithium. 

\begin{figure}
\centering
\subfigure{
\begin{minipage}{\columnwidth}
\includegraphics[width=\columnwidth]{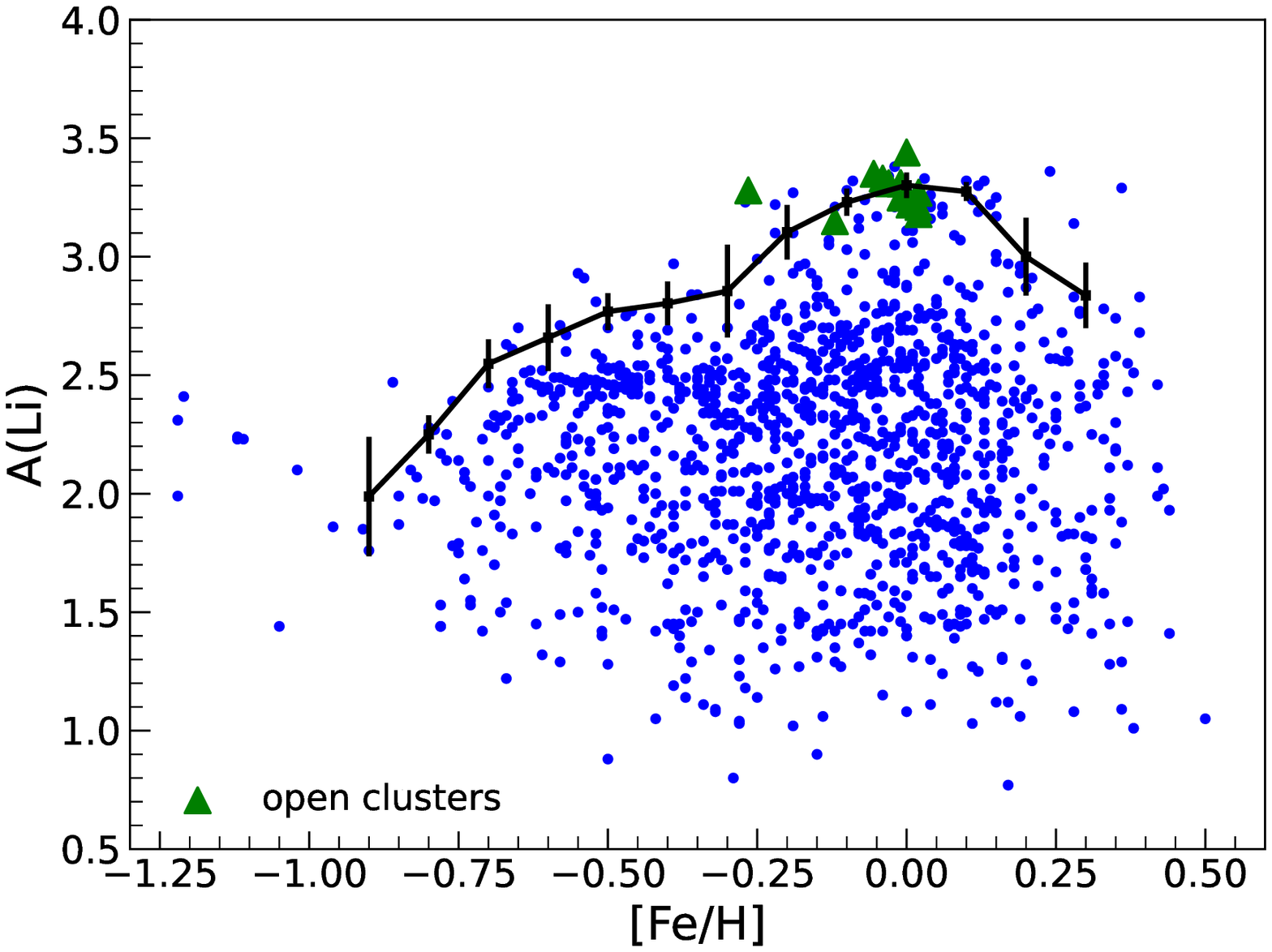}
\end{minipage}
}
\subfigure{
\begin{minipage}{\columnwidth}
\includegraphics[width=\columnwidth]{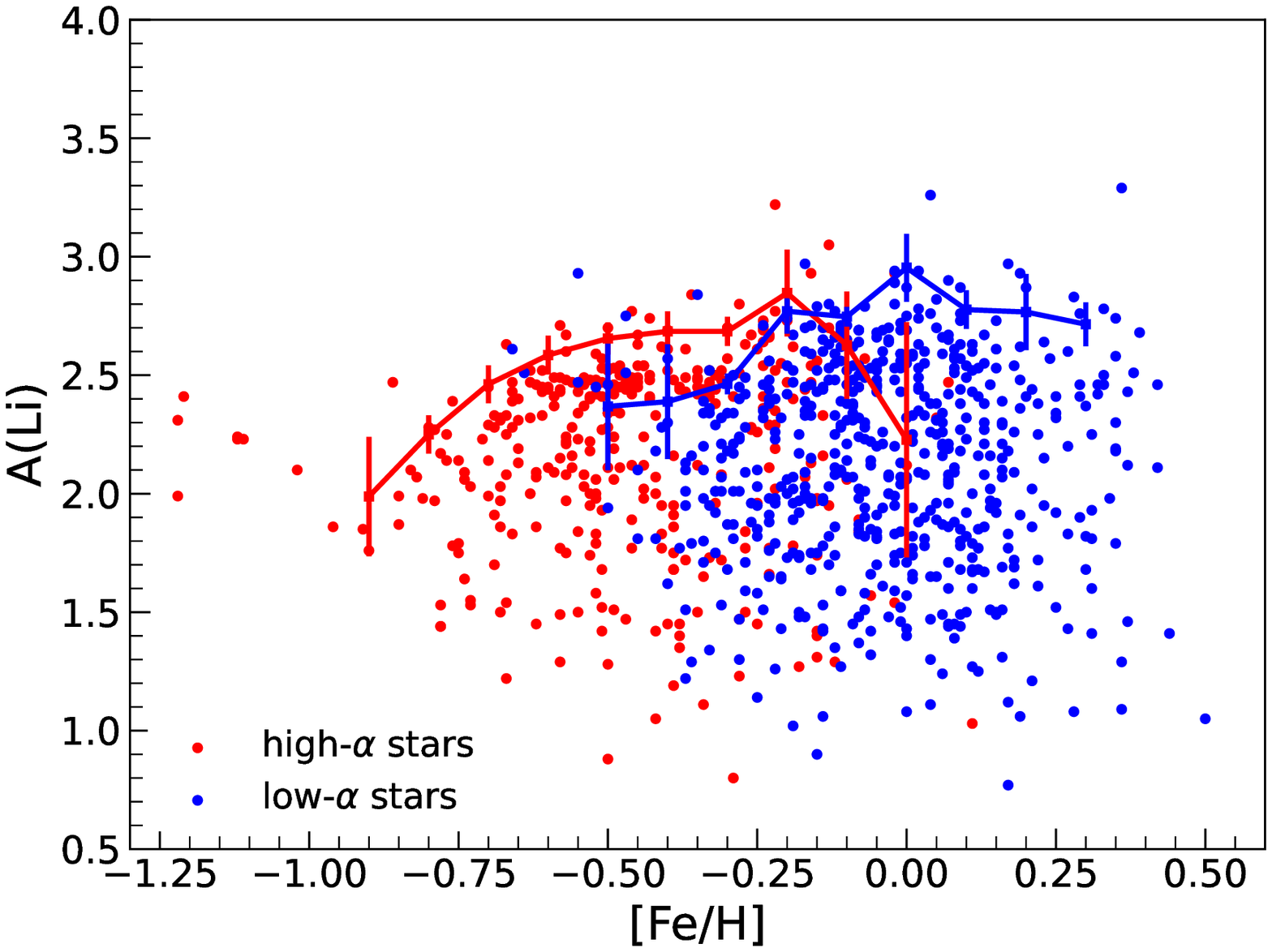}
\end{minipage}
}
\caption{
Upper panel: the lithium abundance vs $\feh$ for open clusters and field stars in the solar neighborhood  (7 kpc $< R_{gc} <$ 9 kpc) of our sample based on \protect\cite{romano2021gaia} (GES iDR6). The solid black line indicates the maximum lithium abundance. Lower panel: same as the upper panel, but for high- (red dots) and low-alpha (blue dots) field stars. The maximum lithium abundance of these two populations is also indicated by solid lines of the same color as each population. See text for details.} \label{fig:1}
\end{figure}

\subsection{Lithium upper envelopes at different $R_{birth}$}
Although most of the stars in the sample are located in the solar neighborhood, their migration distance suggest that many of them may have migrated to the solar neighborhood from the inner or outer, as shown in the upper panel of Fig.~2.
We also plot A(Li) for open clusters and the maximum lithium abundance for field stars both with the absolute values of migration distances less than 1 kpc, as shown in the lower panel of Fig.~2. 
For in-situ stars, the maximum lithium abundance calculated using the same method is still decreasing at the metal-rich side, either because it is the different regions of the disk that have different lithium abundance evolution, or because the metal-rich stars are located in the inner disk and are slightly older.
Meanwhile, if the upper envelope calculated from the open clusters or the 2 field stars with the highest lithium abundance is taken, the maximum lithium abundance is always increasing.
This result suggests that the use of young stars is the better way to track the evolution of lithium.

\begin{figure}
\centering
\subfigure{
\begin{minipage}{\columnwidth}
\includegraphics[width=\columnwidth]{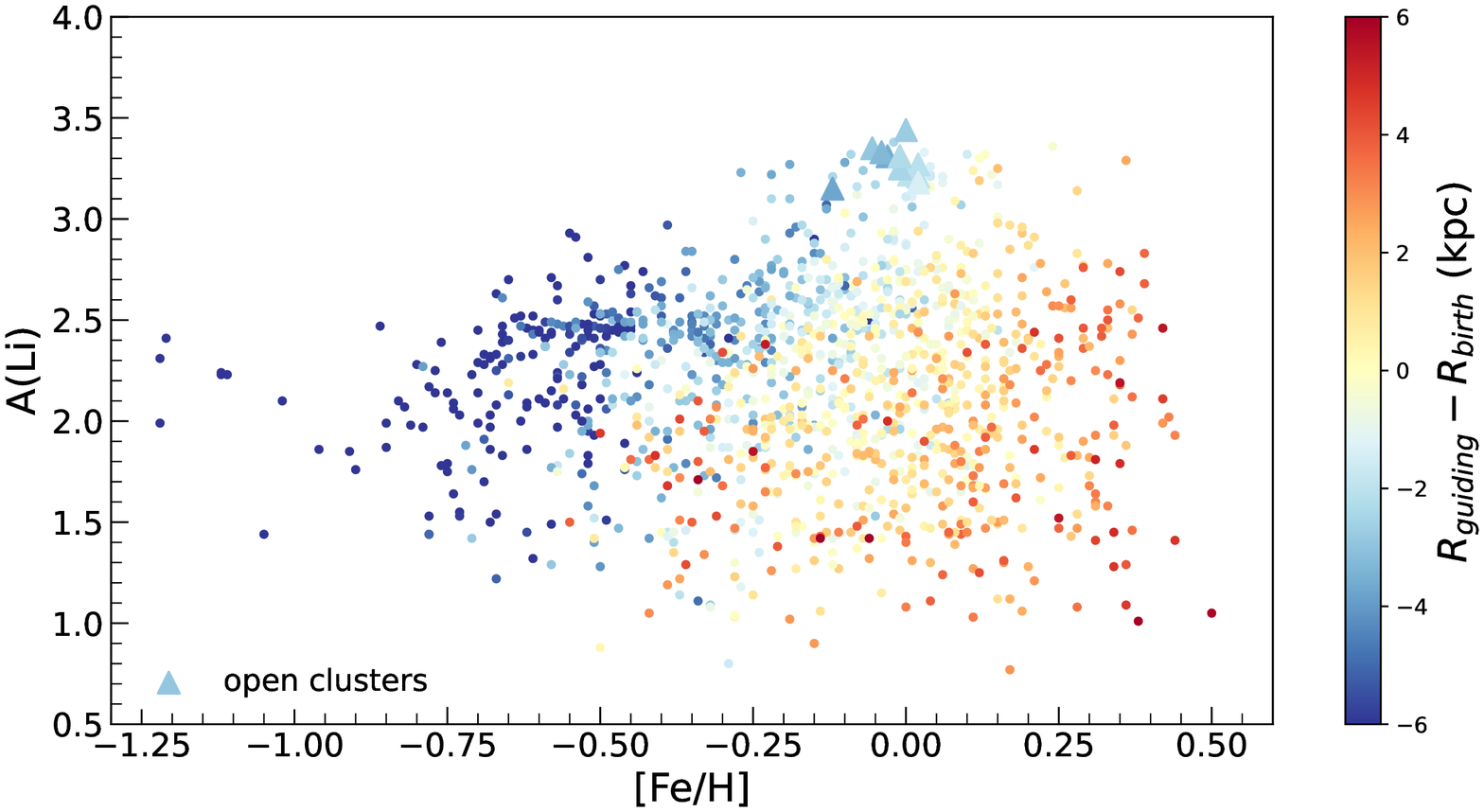}
\end{minipage}
}
\subfigure{
\begin{minipage}{\columnwidth}
\includegraphics[width=\columnwidth]{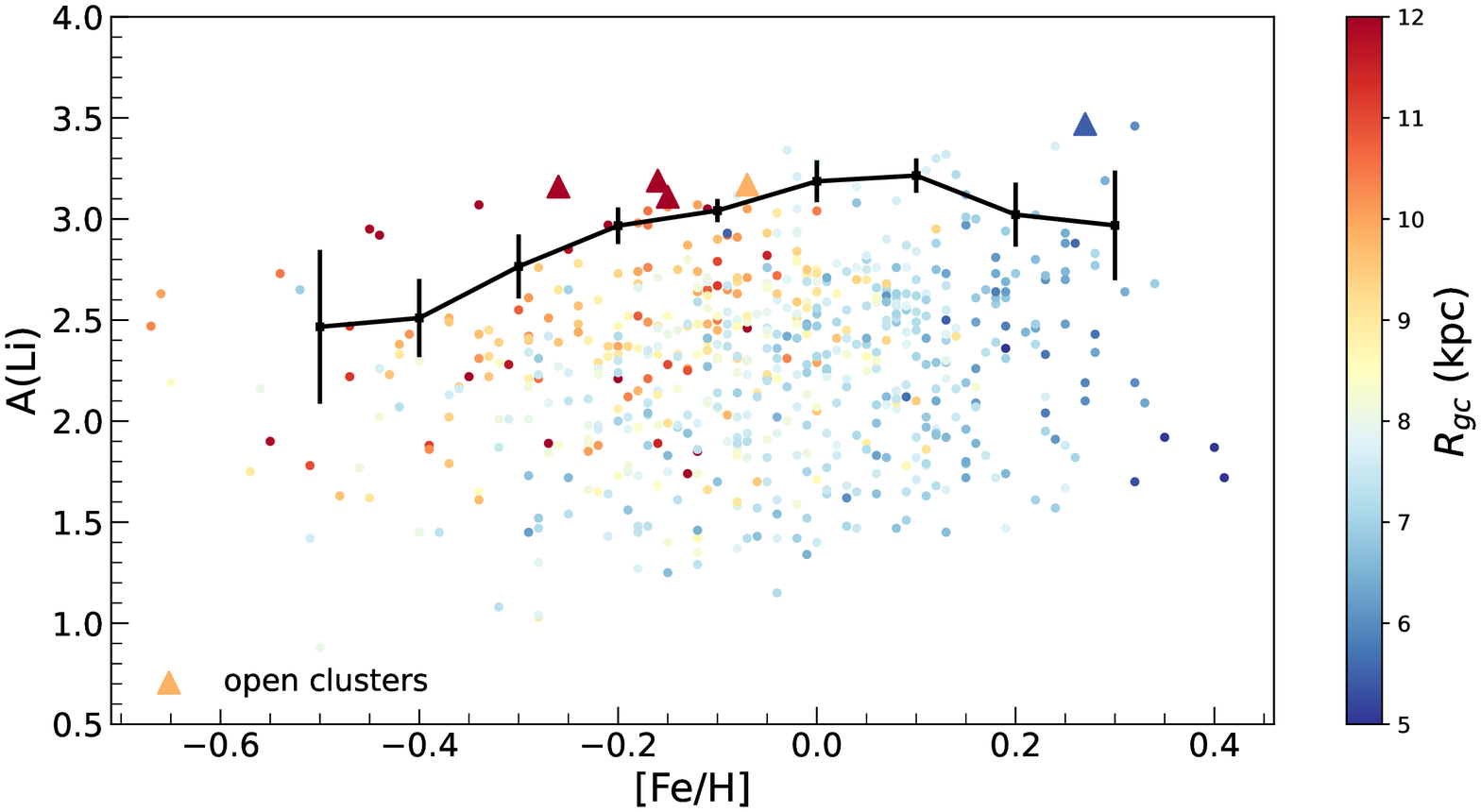}
\end{minipage}
}
\caption{
Upper panel: same as the upper panel in Fig.~1, but the colors indicate their migration distances. Lower panel: A(Li) for open clusters and the maximum lithium abundance for open clusters both with the absolute values of migration distances less than 1 kpc (in-situ stars), with colors indicating $R_{gc}$. The method for calculating the maximum lithium abundance line is the same as in Fig.~1.} \label{fig:2}
\end{figure}

The evolution of lithium abundance with metallicity was different at different galactocentric distances and therefore, we analysed the variation of the maximum lithium abundance with metallicity by binning the stars in the sample according to $R_{birth}$ rather than $R_{gc}$.
The upper panel of Fig.~3 shows the maximum lithium abundance as a function of $\feh$ for six different $R_{birth}$ bins. 
Since the number of stars is lower for 1 $< R_{birth} <$ 3 kpc and 3 $< R_{birth} <$ 5 kpc, the width of bins must be expanded to ensure that there are enough stars ($\geq 6$) in each bin to calculate the maximum lithium abundance, so there are fewer points on this figure.
As can be seen in this figure, stars born in different regions are distributed in different metallicity ranges: those born in the bar/bulge region have highest average metallicities, while those born in the region larger than 11 kpc have the lowest average metallicities. 
Accordingly, for the total sample (see black solid line in the upper panel of Fig.~3), the birth positions of stars at different $\feh$ differ.
We found that for each $R_{birth}$ bin the average maximum lithium abundance increased with $\feh$, and the average maximum lithium abundance is higher at the same $\feh$ with a larger $R_{birth}$.
These results are similar to the right panel of Fig.~7 of \cite{minchev2019yule}, in that they also studied the variation of lithium abundance with metallicity in mono-$R_{birth}$ populations, although they used the mean value of A(Li) rather than the maximum lithium abundance.
According to inside-out chemical evolution models \citep[e.g.][]{chiappini2001abundance}, the ISM of the inner disk is enriched to the same metallicity earlier than that of the outer disk, so that stars with the same metallicity are younger in the outer disk than in the inner disk.
And the lithium decreases with age as a result of stellar evolution, so that for stars of the same metallicity, the lithium depletion of the stars born in the outer disk is less.
At $\feh >$ 0.1 dex, the stars in the sample have $R_{birth}$ basically less than 7 kpc, i.e., they are born within the solar orbit.
The average $R_{birth}$ of these stars is 5.2 kpc, the average migration distance is 1.2 kpc, and the average age is 4.7 Gyr.
Since radial migration of stars needs time and young stars born in this region do not have enough time to migrate to the solar neighborhood to be rarely observed by us \citep[e.g.][]{quillen2018migration,chen2020open}, the envelope is not representative of the undepleted lithium abundance, thus causing the decrease in lithium abundance of the total sample at super-solar metallicities.

We also show the average maximum lithium abundance of low-$\alpha$ stars at different $R_{birth}$ in the lower panel of Fig.~3. 
It can be seen that the distribution trends for each $R_{birth}$ bin of the low-$\alpha$ stars are the same as the upper panel, except for stars with $R_{birth} >$ 11 kpc which do not give an envelope and curves for 5 $< R_{birth} <$ 7 kpc and 7 $< R_{birth} <$ 9 kpc are slightly intersected due to the lack of enough stars.
The lithium abundances of the envelope are smaller than those of the total sample (upper panel) at the same $\feh$, due to the lack of $\alpha$ abundances in the catalog for the stars with the highest lithium abundances.

Although the effect of radial migration on the thick disk is not yet clear \citep[e.g.][]{schonrich2009chemical,minchev2012radial,schonrich2017understanding}, \cite{romano2021gaia} indicated that only 25\% of the high-$\alpha$ stars are thick-disk stars if we distinguish between thick- and thin-disk stars in the sample according to kinematic criteria (total velocities with respect to the LSR between 85 - 210 km/s), and none of them are the highest lithium abundance stars. 
If age selection was added, there were no thick disc stars with A (Li) $>$ 2.5 dex. 
The comparison of the upper and lower panels in Fig.~3 also show the reliability of the findings of the above analysis.
Therefore, we can safely study the effect of radial migration on the lithium abundance distribution using the total sample without having to discard a large number of stars that lack $\alpha$ abundance information.

\begin{figure}
\centering
\subfigure{
\begin{minipage}{\columnwidth}
\includegraphics[width=\columnwidth]{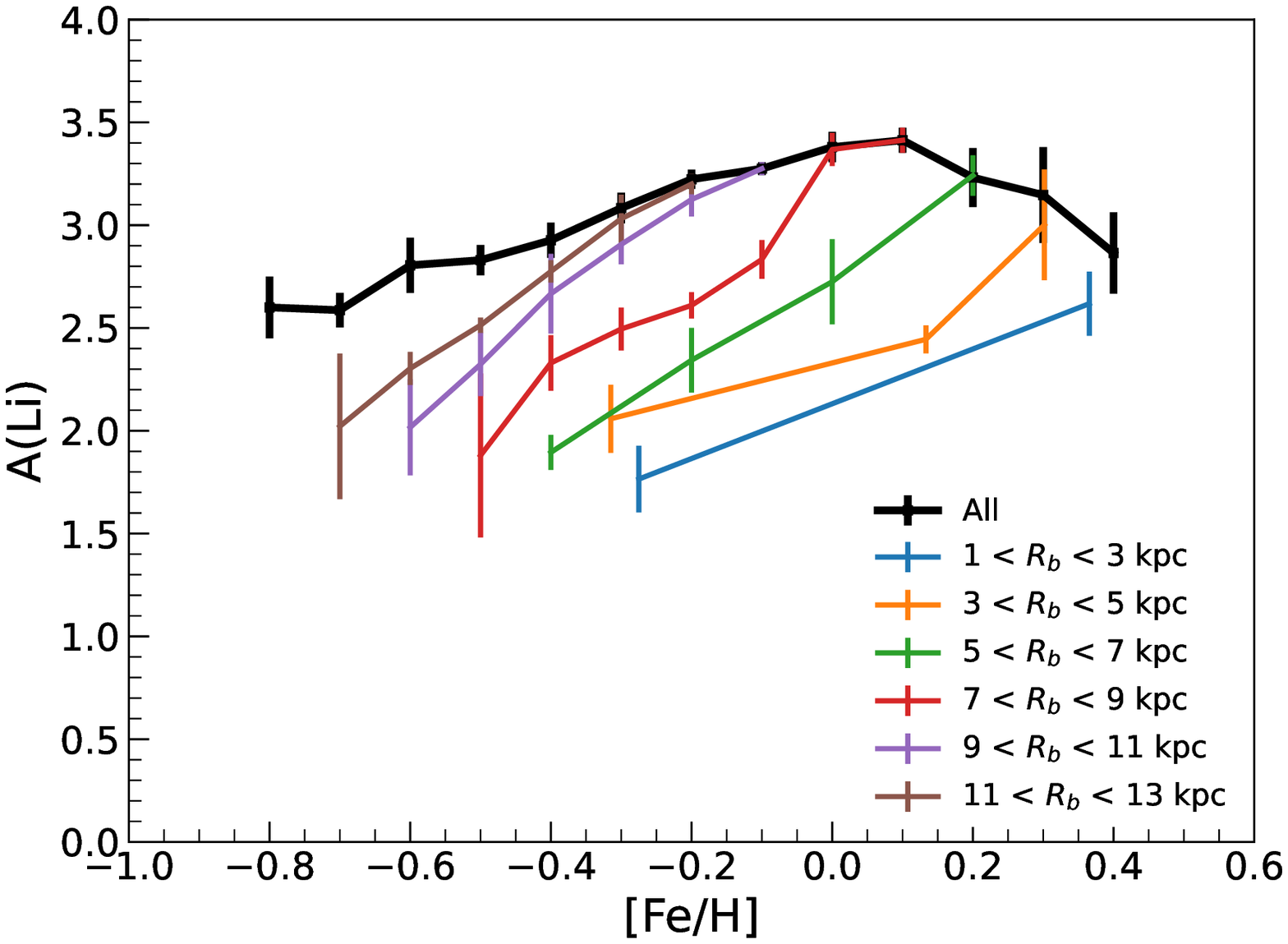}
\end{minipage}
}
\subfigure{
\begin{minipage}{\columnwidth}
\includegraphics[width=\columnwidth]{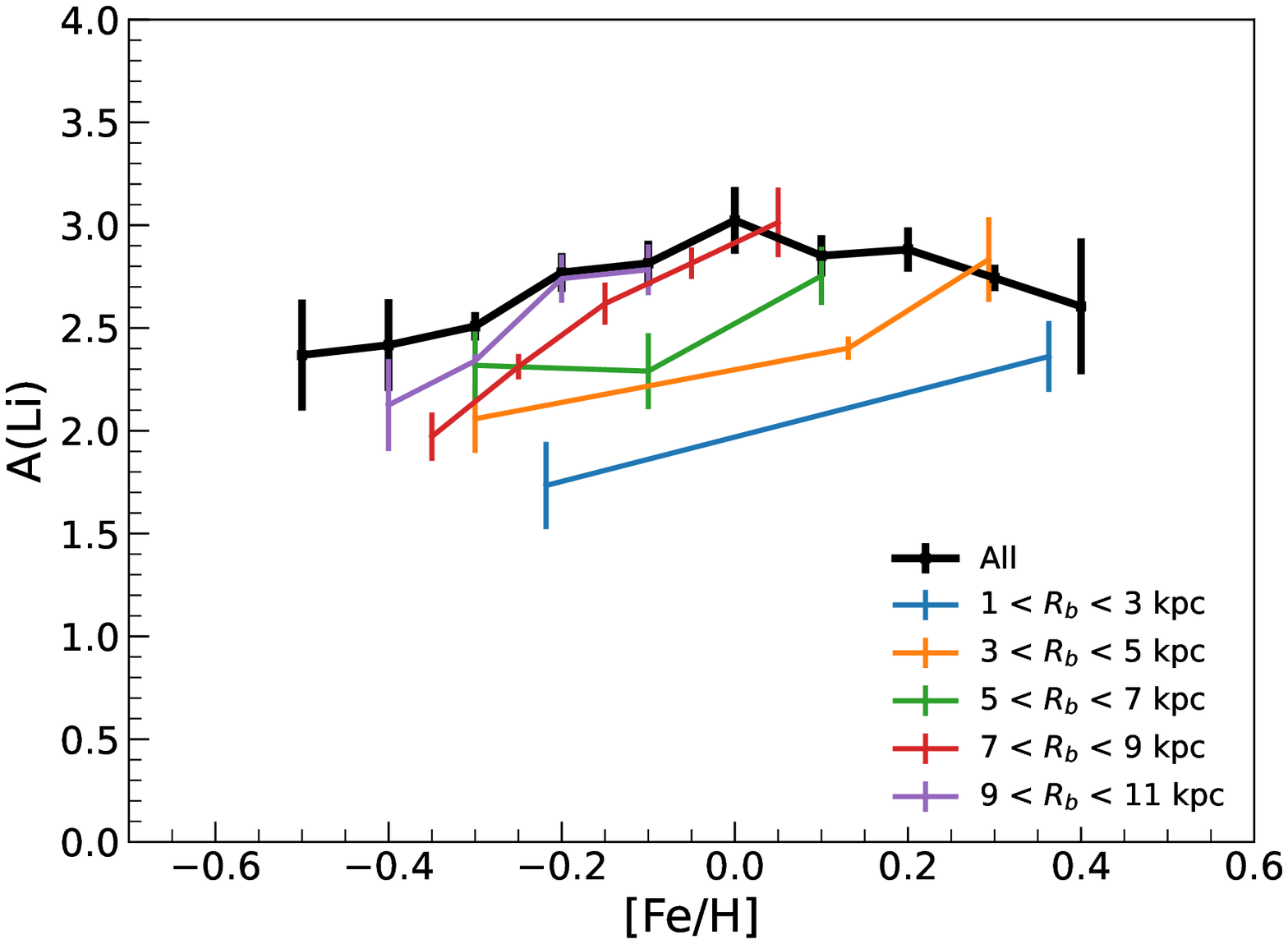}
\end{minipage}
}
\caption{
Upper panel: the maximum lithium abundance of field stars for different $R_{birth}$ bins, with that of the total sample overplotted on the figure with a solid black line. The method for calculating maximum lithium abundance lines is the same as in Fig.~1. Lower panel: same as the upper panel, but for low-$\alpha$ stars.} \label{fig:3}
\end{figure}

\subsection{Lithium abundance as a function of $R_{birth}$}
In this section, we focus on the evolution of the Galactic lithium gradient after excluding the effect of radial migration.
Since radial migration relocates stars from their birth position to their current position, we believe it is more reasonable to use the birth position $R_{birth}$ of the star as the radial distance scale to study the evolution of the ISM than the current position $R_{gc}$, which excludes the influence of radial migration.
The upper panel of Fig.~4 shows the comparison of A(Li) of open clusters and the maximum lithium abundance of field stars with the theoretical gradients.
We removed the open clusters that could not give accurate metallicities, because this would significantly affect the accuracy of the calculated $R_{birth}$.
We calculated the maximum lithium abundance of the field stars in the same method as in Fig.~1, except that the $\feh$ bin is replaced by the $R_{birth}$ bin. 
The red line is derived using the minimum age of the field stars of 1 Gyr.
To easily compare with the lithium abundance distribution scaled by the current position of the stars, $R_{gc}$ \citep[Fig.~7 and 8 in][]{romano2021gaia}, we used the same theoretical model gradient.
The fiducial model is a 'parallel Galactic chemical evolution (GCE) model' by \cite{grisoni2017ambre,grisoni2018abundance}, specifically using the prescriptions of model 1 IM B of \cite{grisoni2018abundance}, later calibrated by observational data, with the SBBN-predicted value used for the primordial lithium abundance and without considering the contribution of the $\nu$-process to the lithium abundance.
The alternative model reduces the mass range of the primary stars entering the formation of nova systems (minimum mass from 1.0 to 3.0 \(\textup{M}_\odot\)) and the nova outburst rate \citep[the total amount of $^7$Li ejected during the lifetime of a typical nova from $2.55 \times 10^{-6}$ to $1.45 \times 10^{-6}$ \(\textup{M}_\odot\), see][for details]{grisoni2019evolution,romano2021gaia}.

In the upper panel of Fig.~4, it can be seen that the A(Li) of open clusters and the maximum lithium abundance of field stars lie between the present fiducial model gradients and alternative model gradients, being closer to both at 8 $\leq R_{birth} \leq 11$ kpc, but closer to the alternative model gradients at $R_{birth} =$ 7 and 12 kpc.
At $R_{birth} <$ 7 and $>$ 12 kpc, the maximum lithium abundance differs significantly from the fiducial and alternative model gradient.
However, since the alternative model gives a flatter gradient in the inner disk, open clusters and field stars with the highest lithium abundance at $R_{birth} =$ $5 - 6$ kpc are closer to the alternative model values and significantly different from the fiducial model values, which suggests that the alternative model is more reliable.
According to the average rate of radial migration of 1 kpc/Gyr \citep{quillen2018migration}, stars born far from the solar neighborhood would take several Gyr to reach the solar neighborhood, so the sample lacks such stars.
Also, in the middle panel of Fig.~4, it can be seen that the difference between the maximum lithium abundance of stars with ages $>$ 1 Gyr and $>$ 4.5 Gyr decreases in this region, again due to the lack of young stars.
At 7 $\leq R_{birth} \leq$ 12 kpc with sufficient number of stars, the difference between the upper envelope of stars with ages $>$ 4.5 Gyr and the dashed gray line is around 0.3 dex, which indicates that the maximum lithium abundance depleted by around 0.3 dex during this time.
The lower panel of Fig.~4 shows the age distribution used to calculate the maximum lithium abundance stars in each $R_{birth}$ bin.
Only when there are sufficient stars with ages less than 3 Gyr in the bin, the calculated present maximum lithium abundance would approach the theoretical value.

At $R_{birth} >$ 12 kpc, no trend of flattening of the A(Li) gradient was found, nor a similar situation as for \cite{romano2021gaia} where the lithium abundance exceeded the theoretical value.
This phenomenon indicates that the observational data cannot prove that the models underestimate the present-day lithium abundance in the outer disk, and that the lack of stars, especially young ones, is the main reason for the lower maximum lithium abundance of stars than the theoretical value.

\begin{figure}
\centering
\subfigure{
\begin{minipage}{\columnwidth}
\includegraphics[width=\columnwidth]{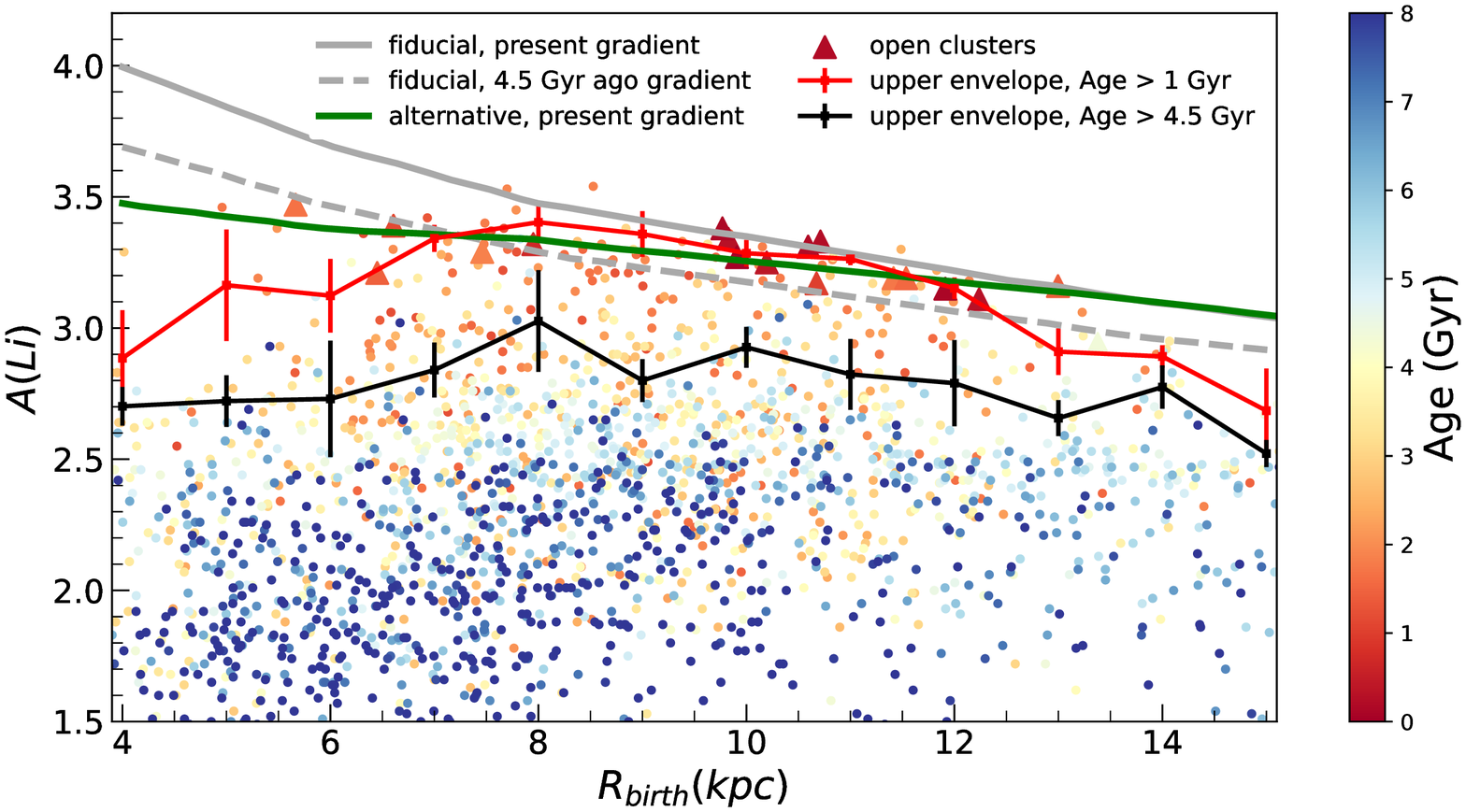}
\end{minipage}
}
\subfigure{
\begin{minipage}{\columnwidth}
\includegraphics[width=\columnwidth]{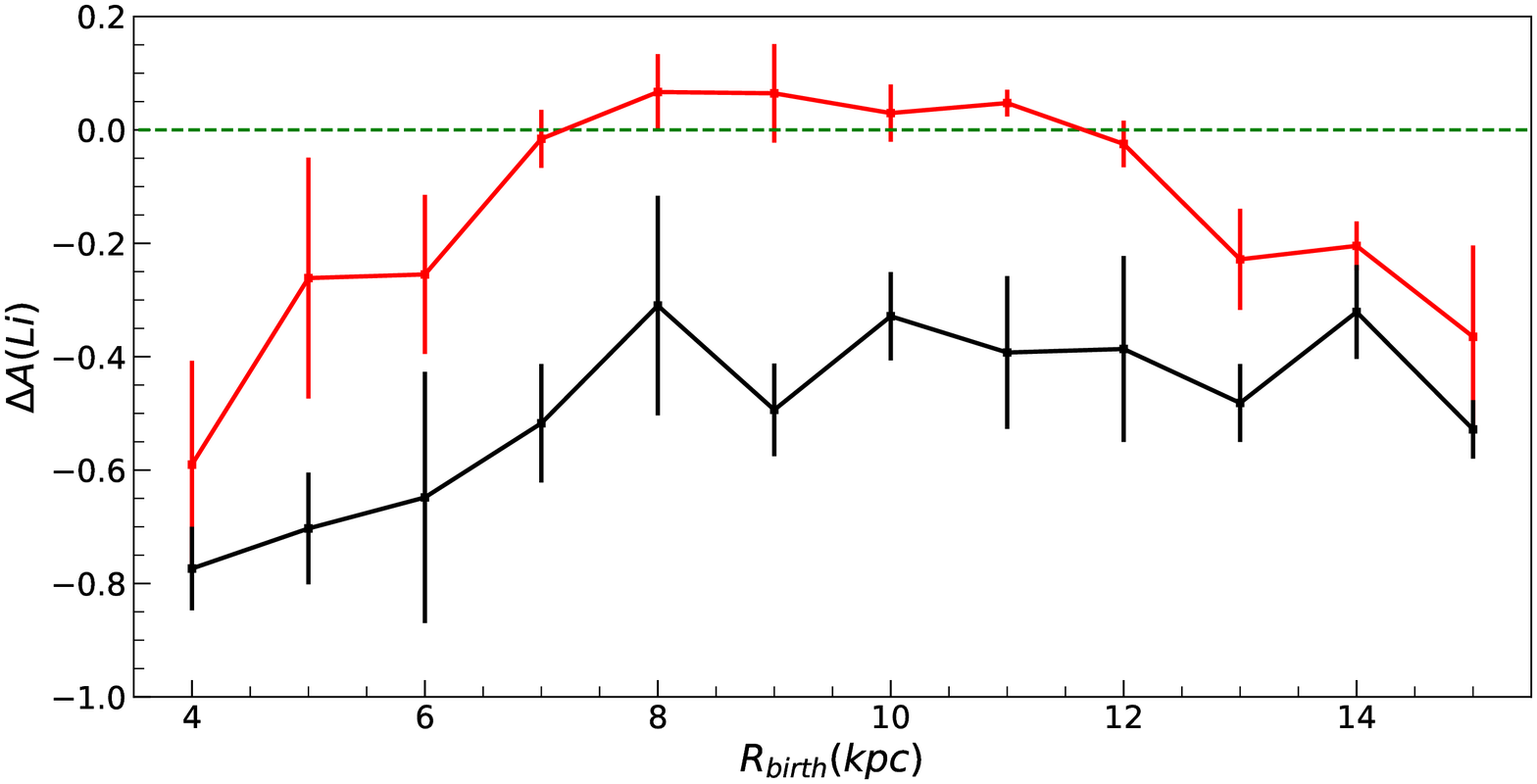}
\end{minipage}
}
\subfigure{
\begin{minipage}{\columnwidth}
\includegraphics[width=\columnwidth]{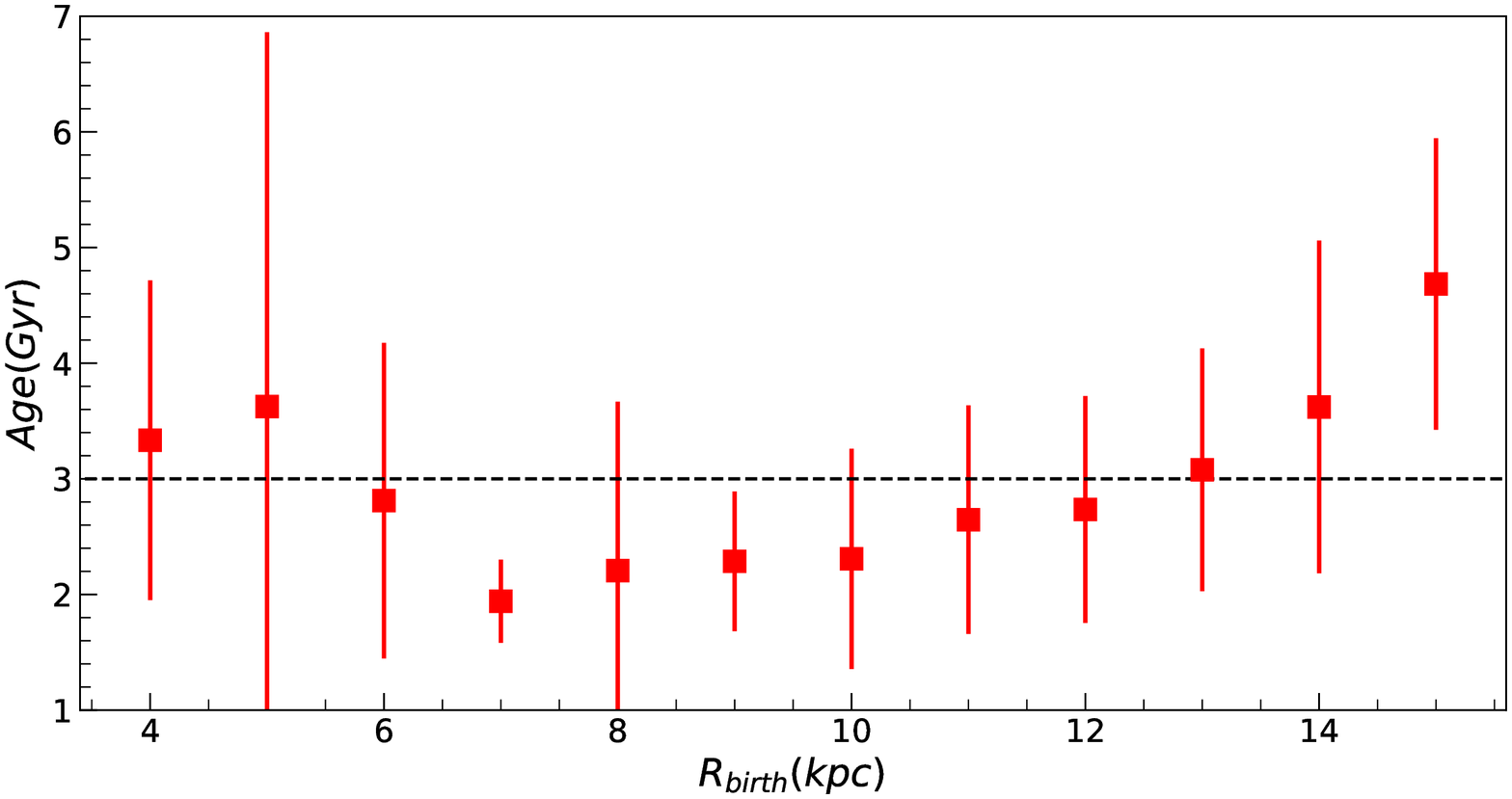}
\end{minipage}
}
\caption{
Upper panel: radial distributions of the lithium abundance of open clusters with accurate $R_{birth}$ and all field stars, using $R_{birth}$ as the radial scale and color coded by age. The solid and dashed gray lines represent the present and 4.5 Gyr ago fiducial GCE model gradients, and the solid green lines represent the present alternative model gradient, respectively, all of which are described in \protect\cite{romano2021gaia} (see text for details). The red and black lines indicate the maximum lithium abundance of all field stars and age $>$ 4.5 Gyr field stars, respectively. The method for calculating maximum lithium abundance lines is the same as in Fig.~1. Middle panel: the difference between the maximum lithium abundance and the present alternative model gradient in the upper panel. Lower panel: the average age of the six stars used in each $R_{birth}$ bin for the calculation of the maximum lithium abundance of all field stars in the upper panel, with the error bars indicating the standard deviation.}
\label{fig:4}
\end{figure}

\subsection{Evolution of lithium abundance in the solar neighbourhood}
Since the initial atmospheric chemical abundances of the stars depend on their surrounding ISM at the time of their birth, we should use the chemical evolution model near their $R_{birth}$ in comparison to the model near their present location $R_{gc}$.
Here, we selected stars with $6 < R_{birth} < 10$ kpc to compare with model predictions for the following reasons: (i) open clusters and field stars near the Sun; (ii) this region has a high number of field stars, accounting for 40\% of the total sample, as well as a high number of young stars, which is statistically significant; (iii) the lithium abundances of clusters and maximum lithium abundances in this region fit well with the alternative model, except at $R_{birth} =$ 6 kpc; (iv) the difference in A(Li) predicted by the alternative model between 6 and 10 kpc is very small, so the effect of $R_{birth}$ need not be considered in the following analysis.

Fig.~5 shows the average maximum lithium abundance of stars born in the vicinity of the sun. 
With increased metallicity, the average maximum lithium abundance increases from 2.0 dex at $\feh =$ -0.65 to 3.4 dex at $\feh =$ 0.15 dex, with no downward trend in lithium abundance.
Note that the A(Li) of the stars do not exceed any of the theoretical values at $\feh =$ -0.5 to -0.1 dex, which differs from Fig.~8 of \cite{romano2021gaia}.
This can be explained by the fact that the lithium abundance of this part of stars was depleted with time and thus lower than the theoretical value, without modifying the model.
However, at $\feh =$ -0.05 to 0.15 dex, the maximum lithium abundance is similar to the alternative model, but is slightly larger by 0.15 dex, and may require subsequent fine-tuning of the model for the evolution of lithium abundance and metallicity with time.
This range of metallicities corresponds to the age less than 3 Gyr on the lower panel of Fig.~5, and stars older than this age have increasing gaps with the model predictions because the maximum lithium abundances no longer represent the initial value due to depletion.
At the same time, we can see that the lithium abundances of these young open clusters fit well with the alternative model.

\begin{figure}
\centering
\subfigure{
\begin{minipage}{\columnwidth}
\includegraphics[width=\columnwidth]{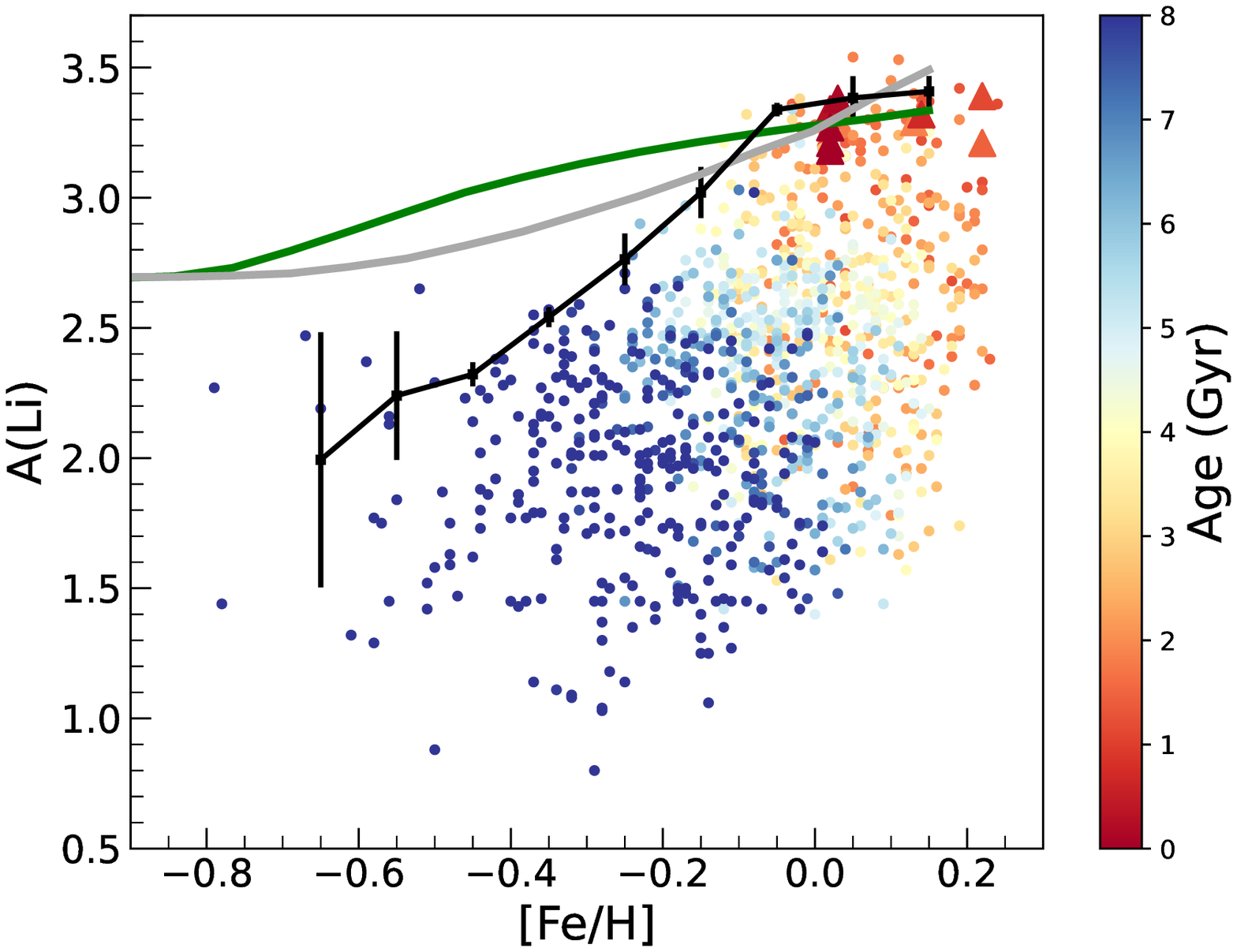}
\end{minipage}
}
\subfigure{
\begin{minipage}{\columnwidth}
\includegraphics[width=\columnwidth]{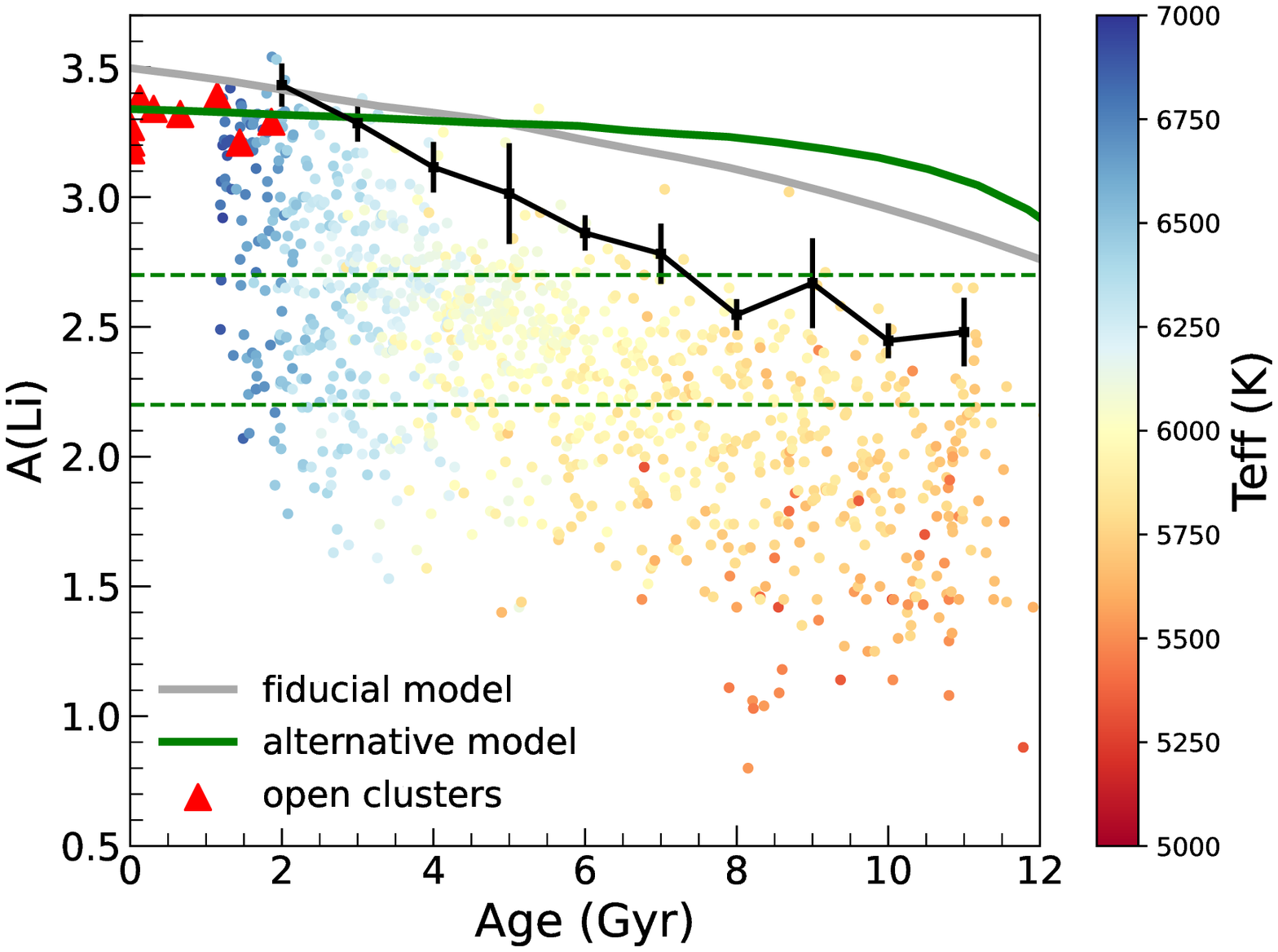}
\end{minipage}
}
\caption{
Lithium abundance vs $\feh$ and Age for open cluster and field stars born in the solar neighborhood ($6 < R_{birth} < 10$ kpc). The solid black line is the average maximum lithium abundance, using the same method as in Fig.~1. The solid gray and green lines respectively indicate the predictions of the fiducial and alternative models described in \protect\cite{romano2021gaia}. The upper and lower horizontal green dashed lines in the lower panel represent the primordial lithium abundances predicted by SBBN \protect\citep{pitrou2018precision} and 'Spite plateau' \protect\citep{bonifacio1997primordial}, respectively.}
\label{fig:5}
\end{figure}

\section{Conclusions}
Based on the lithium abundance catalog from \cite{romano2021gaia} and the ISM metallicity evolution model from \cite{minchev2018estimating}, we calculated $R_{guiding}$ and $R_{birth}$ for open clusters and field stars to analyze the role of radial migration on the evolution of the lithium abundance in the disk of the Milky Way.

Using the maximum lithium abundance envelope of the field stars, we find that the lithium abundance starts to decrease at $\feh =$ 0 dex in the solar neighborhood (7 $< R_{gc} <$ 9 kpc) and at 0.1 dex in the total sample (mainly 5 $< R_{gc} <$ 12 kpc).
The stars in the sample with $\feh$ greater than 0.1 dex have $R_{birth}$ basically less than 7 kpc, with an average of 5.2 kpc, an average outward migration of 1.2 kpc, and an average age of 4.7 Gyr, indicating that most of them are intermediate-age stars born in the inner disk migrating outward to the solar neighborhood.
For stars with the same $R_{birth}$, their maximum lithium abundance does not show a decrease at the metal-rich end, suggesting that the decrease in lithium abundance is not a real decrease during chemical evolution, but a facade due to radial migration.

In the A(Li)-$R_{birth}$ panel, we compare A(Li) of open clusters and the maximum lithium abundance of field stars with the model gradients of the fiducial GCE model and the alternative model \citep{grisoni2017ambre,grisoni2018abundance,romano2021gaia} and find that the latter fit the observations well at $R_{birth} =$ 7 - 12 kpc.
The better fit of the model gradient to observations on the A(Li)-$R_{birth}$ panel than on the A(Li)-$R_{gc}$ panel confirms the necessity of considering radial migration when studying the evolution of the lithium abundance.
We also find that based on stars with ages less than 3 Gyr, the maximum lithium abundance obtained is closer to the undepleted value.
When the stars are of solar age (around 4.5 Gyr), the maximum lithium abundance depletes by around 0.3 dex from the original value.

By comparing the sample with the theoretical model, we find that stars born in the solar neighborhood do not show significant lithium depletion at $\feh >$ -0.05 dex or age $<$ 3 Gyr. 
This result provides a limit on the metallicity and age to study the evolution of lithium abundance in the future.

\section*{Acknowledgements}

This study is supported by the National Natural Science Foundation of China under grant Nos. 11988101, 11890694, 11873052, the CSST project, the National Key R\&D Program of China No. 2019YFA0405500, CAS Project for Young Scientists in Basic Research, Grant No. YSBR-062 and the Joint Research Fund in Astronomy (U2031203) under cooperative agreement between the National Natural Science Foundation of China (NSFC) and Chinese Academy of Sciences (CAS). 

\section*{Data Availability}
Data and models used in this paper are available from the corresponding papers and the $R_{guiding}$ and $R_{birth}$ calculated in this paper are available through the authors.



\bibliographystyle{mnras}
\bibliography{Li} 

\bsp	
\label{lastpage}
\end{document}